\documentclass[preprint,5p,times,twocolumn]{elsarticle}

\usepackage{amssymb}
\usepackage{amsmath}

\usepackage{amsfonts}
\usepackage{algorithmic}
\usepackage{graphicx}
\usepackage{textcomp}
\usepackage{xcolor}

\usepackage{verbatim, booktabs, multirow}
\usepackage[marginal]{footmisc}

\journal{Speech Communication}

\begin{document}

\begin{frontmatter}



\title{Trusted Fake Audio Detection Based on Dirichlet Distribution}


\author[1,2]{Chi Ding}
\author[1]{Junxiao Xue \corref{cor1}}

\author[1]{Cong Wang \corref{cor1}}
\author[3]{Hao Zhou}
\cortext[cor1]{Corresponding author. \\
E-mail address: xuejx@zhejianglab.com.\\
E-mail address: cong.wang@zhejianglab.com.}
\affiliation[1]{organization={Research Center for Space Computing System, Zhejiang Lab}, 
            city={Hangzhou},
            country={China}
            }
\affiliation[2]{organization={Hangzhou Institute for Advanced Study, University of Chinese Academy of Sciences}, 
            city={Hangzhou},
            country={China}
            }
\affiliation[3]{organization={School of Computer Science and Engineering, National Engineering Research Center of Digital Life, Sun Yat-sen University}, 
            city={Guangzhou},
            country={China}
            }

\begin{abstract}
With the continuous development of deep learning-based speech conversion and speech synthesis technologies, the cybersecurity problem posed by fake audio has become increasingly serious. Previously proposed models for defending against fake audio have attained remarkable performance. However, they all fall short in modeling the trustworthiness of the decisions made by the models themselves. Based on this, we put forward a plausible fake audio detection approach based on the Dirichlet distribution with the aim of enhancing the reliability of fake audio detection. Specifically, we first generate evidence through a neural network. Uncertainty is then modeled using the Dirichlet distribution. 
By modeling the belief distribution with the parameters of the Dirichlet distribution, an estimate of uncertainty can be obtained for each decision. Finally, the predicted probabilities and corresponding uncertainty estimates are combined to form the final opinion. On the ASVspoof series dataset (i.e., ASVspoof 2019 LA, ASVspoof 2021 LA, and DF), we conduct a number of comparison experiments to verify the excellent performance of the proposed model in terms of accuracy, robustness, and trustworthiness.
\end{abstract}




\end{frontmatter}




\section{Introduction}
\label{sec1}

The swift advancement of information technology has led to notable progress in speech synthesis and speech conversion technologies, enabling the effortless generation of high-quality speech \cite{ling2015deep, min2021meta, theune2001data}. This rapid evolution is driven by breakthroughs in deep learning, particularly generative models such as Generative Adversarial Networks (GANs) \cite{kong2020hifi,gogate2024robust}, Variational Autoencoders (VAEs) \cite{xiao2024eggesture}, and diffusion model \cite{evans2024fast}, which allow for increasingly realistic and human-like synthetic speech. However, this progress has also given rise to cybersecurity concerns related to fake audio. Fake audio refers to artificially synthesized or converted voice samples designed to deceive speech recognition systems by incorrectly identifying them as authentic speech. The proliferation of fake audio poses a grave threat to critical domains \cite{chesney2019deepfakes}, including speech recognition, authentication, and security monitoring. Furthermore, the ease of spreading fake audio exacerbates social risks, amplifying the potential for harm across both digital and physical environments.

To tackle the security challenges arising from fake audio, researchers have delved into the field of fake audio detection. Over recent years, researchers have put forth various fake audio detection methodologies \cite{borrelli2021synthetic, lv2022fake, wu2022partially, ma2021continual,xue2023cross}, encompassing acoustic feature-based approaches, convolutional neural network-based techniques, and transformer-based methods. Despite they have achieved impressive performance, current detection models often fail to accurately quantify their own confidence levels. This limitation is particularly problematic in critical applications where incorrect decisions could have severe consequences. 

The assessment of decision uncertainty is critical in real-world applications. By leveraging model confidence, we can effectively address uncertain samples and specific situations.
For example, if a fake audio detection model returns a highly uncertain classification result, the input can be forwarded to human experts for manual review or to a more advanced model for further analysis. This helps avoid erroneous decisions and improves the model’s reliability and accuracy.
Moreover, in certain scenarios such as medical diagnosis and financial risk assessment, decision uncertainty evaluation becomes particularly important, as it helps identify potential risks and take appropriate measures to mitigate them. Therefore, evaluating model reliability is one of the key steps in building trustworthy detection systems.

However, it is a common challenge that standard deep learning models often struggle to capture prediction uncertainty\cite{gawlikowski2023survey}. This issue stems from the conventional training paradigm, where networks are optimized solely to minimize prediction loss. As a result, the trained models focus on maximizing accuracy but remain unaware of their own confidence in the predictions, leading to overconfident outputs even when faced with ambiguous or adversarial inputs. In classification tasks, the predicted probabilities obtained at the end of the pipeline are often misinterpreted as model confidence, but even with high outputs, the model's predictions can still exhibit uncertainty\cite{2015Dropout}. Researchers have carried out a lot of work to deal with it, including Bayesian Neural Networks (BNNs) \cite{wilson2020bayesian}, Ensemble methods\cite{lakshminarayanan2017simple}, and Evidential neural networks\cite{sensoy2018evidential}. 
Among them, Evidential neural networks can provide stable and high-quality uncertainty estimation for classification tasks and demonstrate robustness when facing adversarial samples.

Based on this, we introduce a plausible fake audio detection method based on the Dirichlet distribution. Our method utilizes the original high-performance detection model as an evidential network and cleverly employs the Dirichlet distribution to generate stable and reasonable uncertainty estimates for classification decisions, thereby ensuring the reliability and robustness of fake audio detection.
Specifically, we start by coordinating evidence generation via a neural network. Subsequently, we model the uncertainty using a Dirichlet distribution. By modeling the belief distribution of decisions using the parameters of the Dirichlet distribution, we determined the uncertainty estimates for model prediction. 
Finally, the predicted probabilities for each category with the corresponding uncertainty estimates will be obtained.

To summarize, our major contributions are twofold:

\begin{itemize}

    \item We present a novel approach to fake audio detection that utilizes the Dirichlet distribution to model uncertainty. Our method estimates the uncertainty associated with each decision using the Dirichlet distribution. This enables our model to provide not only predictions but also confidence intervals, enhancing the transparency and reliability of the detection process.
    \item Our approach uses the Dirichlet distribution to quantify the uncertainty of its predictions. This functionality empowers our model to flag uncertain predictions for additional review, thereby improving overall system robustness and reliability.
    \item We conducted extensive comparison experiments on the ASVspoof series datasets (ASVspoof 2019 LA, ASVspoof 2021 LA, and ASVspoof 2021 DF) and demonstrated that our proposed model achieves notable improvements in accuracy, robustness, and reliability. 
    Our model consistently outperforms existing methods in terms of several metrics(EER, min t-DCF, aECE, and PCC), highlighting its effectiveness in detecting fake audio.

\end{itemize}

The paper is organized as follows: Section 2 introduces the relevant studies. Section 3 explains the workflow and theory of the method. The experiments are described in section 4. The conclusion is given in section 5.
\section{Related work}

\begin{figure*}[ht]
	\centering
	\includegraphics[scale=0.95]{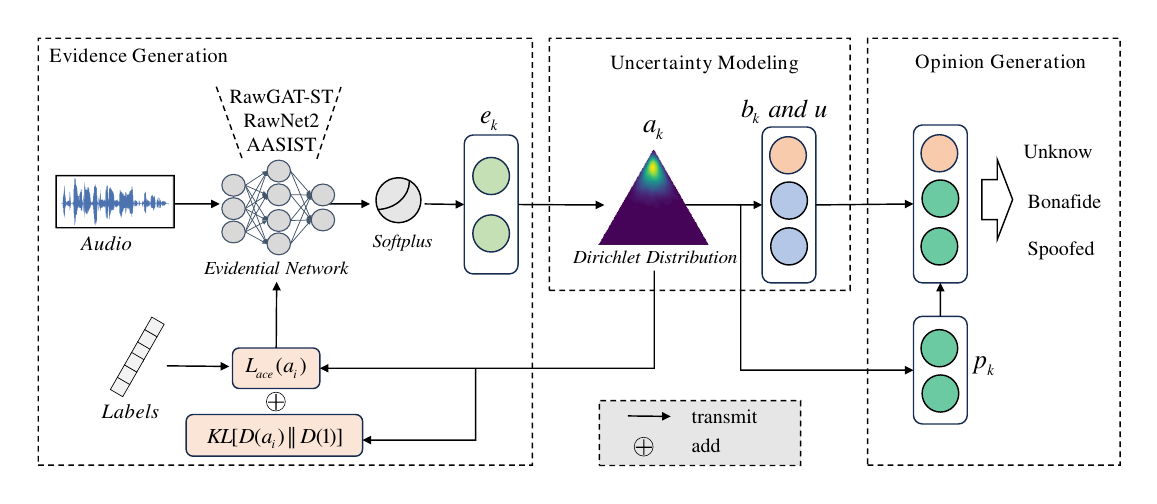}
	\caption{The overall architecture of the proposed trusted fake audio detection method. This method refers to evidence generation, uncertainty modeling based on Dirichlet distribution, and opinion generation as described in Section 2. In the training stage, the evidence is output by the evidential network, and the Dirichlet distribution parameters $a$ determined by evidence are fed into the evidential network to calculate the final loss. In the inference stage, Dirichlet distribution parameters $a$ are used to model belief distribution and form opinion.}
    \label{f1}
\end{figure*}

\subsection{Deep Learning Based Fake Audio Detection Method}
In recent years, deep learning has experienced rapid growth and garnered significant attention both domestically and internationally. Within the field of fake audio detection, an increasing number of researchers have begun exploring and conducting related research using deep learning techniques.

Convolutional neural networks (CNNs) have been widely used in fake audio detection tasks due to their superior ability to capture local spatial correlations. For example, Light CNN (LCNN) \cite{wu2018light} consists of a convolutional layer and a max-pooling layer and employs the Max-Feature-Map (MFM) activation function. LCNN not only demonstrates excellent performance in the LA tasks of ASVspoof 2017 \cite{cheng2019replay} and ASVspoof 2019 \cite{lavrentyeva2019stc}, but its MFM activation function also effectively suppresses environmental noise and signal distortion, thus improving detection robustness.

Although deep CNNs have achieved significant results in spoofed audio detection, the increase in network depth brings problems such as increased training difficulty and performance degradation. To address this challenge, Tomilov et al \cite{tomilov2021stc} and Chen et al \cite{chen2021pindrop} used ResNet as a classifier for deep audio spoofing detection, and achieved excellent results in the ASVspoof 2021 challenge. In addition, Yan et al \cite{yan2022audio} further combined the 34-layer standard ResNet with the multi-attention pooling layer for deep audio detection and won first place in the FG-D task of ADD 2022, which fully demonstrated the excellent performance of the method.
Based on this, Tak et al \cite{tak2021rawnet2} proposed an end-to-end anti-spoofing model, RawNet2, which adopts SincNet \cite{ravanelli2018speaker} as the first layer. SincNet performs the convolution operation by sinusoidal filter and combines with the non-linear transform and the max pooling layer, which realizes the efficient processing of the original waveform and improves the ability to identify fake audio.

As Graph Neural Networks (GNNs) have shown unique advantages in processing complex data structures, researchers have started to explore their applications in false audio detection. Inspired by the success of Graph Attention Network (GAT) \cite{tak2021graph}, Tak et al \cite{tak2021end} proposed a time-frequency graph attention network called RawGAT-ST. The method outperforms the RawNet2 model on the ASVspoof 2019 LA evaluation set by learning the relationships between different audio segments. Subsequently, Jung et al \cite{jung2022aasist} proposed AASIST, a network based on heterogeneous stacked graph attention layers, to model artefacts across time-frequency bands with a heterogeneous attention mechanism, which outperforms the existing state-of-the-art end-to-end models.

Subsequent proposed methods, such as Rawformer \cite{liu2023leveraging}liu, GMM-ResNet2 \cite{lei2024gmm}, and ASSD \cite{liu2025assd}, have achieved better performance. However, these methods focus on classification accuracy but lack in providing confidence in the detection decision. To this end, we address the issue by introducing uncertainty modeling into the detection model.

\subsection{Uncertainty estimation}
In recent years, deep neural networks (DNNs) have achieved remarkable success across various domains, including medical imaging, robotics, and earth observation. However, as these models are increasingly deployed in real-world applications, the reliability of their predictions has become a critical concern. 
Accurate uncertainty estimation is crucial for ensuring the reliability and safety of DNN-based systems. In high-stakes applications such as autonomous driving, healthcare, and financial forecasting, incorrect predictions can have severe consequences. By quantifying uncertainty, practitioners can identify unreliable predictions and take appropriate actions, such as requesting human intervention or gathering additional data. Furthermore, uncertainty-aware models can improve decision-making processes by providing more informative outputs that reflect the level of confidence in each prediction.

Uncertainty can be categorized into two main types: model uncertainty and data uncertainty. Model uncertainty arises when a DNN lacks sufficient information to make confident predictions. It is typically caused by limited training data, inadequate model capacity, or suboptimal hyperparameters. As noted by \cite{guo2017calibration} and \cite{seo2019learning}, deeper networks tend to be more overconfident than shallower ones, highlighting the importance of addressing model uncertainty in complex architectures. Data uncertainty reflects uncertainty caused by variability or noise inherent in the input data. This type of uncertainty cannot be eliminated through better modeling or training but must instead be accounted for in the prediction process. For example, in medical image analysis, data uncertainty may arise from variations in imaging modalities or patient-specific conditions \cite{eaton2018towards}. 

Several approaches have been proposed to estimate uncertainty in DNNs. These methods can be broadly classified into four categories: deterministic neural networks, Bayesian neural networks (BNNs), ensembles of neural networks, and test-time data augmentation approaches.

Deterministic neural networks are the most widely used form of DNNs, where weights are fixed after training. While these networks do not inherently account for uncertainty, several techniques have been developed to approximate both model and data uncertainties using deterministic architectures.
Model uncertainty in deterministic neural networks can be approximated by analyzing the variability in predictions under different conditions. One prominent approach is the Monte Carlo Dropout (MC Dropout), introduced by \cite{gal2016dropout}. Deterministic neural networks typically reflect data uncertainty through a probability distribution of softmax outputs in the classification tasks. BNNs explicitly model uncertainty by treating weights as probability distributions rather than fixed values. Probabilistic backpropagation \cite{hernandez2015probabilistic} and black-box alpha divergence minimization \cite{hernandez2016black} are two prominent techniques for training BNNs. These methods allow for the estimation of both model and data uncertainties, making them particularly suitable for scenarios where reliable uncertainty quantification is essential. Ensemble methods involve training multiple neural networks and aggregating their predictions to estimate uncertainty. With the rise of deep learning, ensemble-based approaches have been extended to uncertainty-aware deep learning, where each member of the ensemble provides a probabilistic output that contributes to the overall uncertainty estimate \cite{lakshminarayanan2017simple}. Test-time data augmentation involves applying transformations to input data during inference to generate multiple predictions. The variability among these predictions can then be used to estimate uncertainty. \cite{wang2019aleatoric} applied this technique in segmentation tasks, demonstrating its effectiveness in capturing pixel-wise uncertainty.

To assess the quality of uncertainty estimates, several metrics have been developed. These metrics evaluate different aspects of uncertainty, such as calibration, sharpness, and coverage probability. Calibration measures the alignment between predicted probabilities and actual outcomes. Specific metrics indicators are Expected Calibration Error (ECE) \cite{guo2017calibration}, adaptive Expected Calibration Error (aECE) \cite{nixon2019measuring}, and so on. 

We choose a deterministic network-based approach to model uncertainty and use aECE to evaluate the quality of uncertainty.

\section{Methodology}

The structure of our method is shown in Figure \ref{f1}. The method is divided into three steps, which are evidence generation, uncertainty modeling, and opinion generation. First, a neural network is used for evidence generation. To satisfy the requirement that the parameters of the Dirichlet distribution must be non-negative, the softmax layer of the generalized neural network is replaced with a non-negative function to obtain an evidence-generating network. Next, uncertainty is modeled using the Dirichlet distribution. By modeling the belief distribution with the parameters of the Dirichlet distribution, an estimate of uncertainty can be obtained for each decision. Finally, the predicted probabilities and corresponding uncertainty estimates for each decision are combined to form a final opinion. 

In this way, a comprehensive assessment of each category is obtained. This combined opinion allows for a more complete understanding of the model's decisions and provides more reliable results. With such a design, it is possible to generate decision opinions with uncertainty estimates, which is important for many application scenarios. 
In summary, the method presented in this section provides a decision-making framework that can integrate the consideration of prediction probability and uncertainty through the steps of evidence generation, uncertainty modeling, and opinion generation. Such a framework can provide a more accurate and reliable assessment of the decision-making process and provide strong support for decision-making in real-world applications.

\subsection{Uncertainty Modeling and Theory of Evidence}

In this subsection, we describe how evidential deep learning can quantify the uncertainty of categorization and how it can model the probability of each category and the overall uncertainty of the current prediction.

Note that our approach focuses on modeling data uncertainty, also referred to as aleatoric uncertainty. In the domain of fake audio detection, this type of uncertainty arises due to various factors such as noise, distortions, or inherent ambiguities in the audio data, which may come from synthetic audio sources or low-quality recordings. 

The Dirichlet distribution is a probability distribution commonly used to model probability vectors. For a probability vector \( \boldsymbol{x} = (x_1, x_2, \dots, x_K) \) representing probabilities of \( K \) categories (where \( \sum_{i=1}^{K} x_i = 1 \) and \( x_i \geq 0 \)), and a parameter vector \( \boldsymbol{\alpha} = (\alpha_1, \alpha_2, \dots, \alpha_K) \) with all \( \alpha_i > 0 \), the Dirichlet distribution's probability density function is:

\begin{equation}
    p(\boldsymbol{x}|\boldsymbol{\alpha}) = \frac{1}{B(\boldsymbol{\alpha})} \prod_{i=1}^{K} x_i^{\alpha_i - 1}
\end{equation}

Where \( B(\boldsymbol{\alpha}) = \frac{\prod_{i=1}^{K} \Gamma(\alpha_i)}{\Gamma(\sum_{i=1}^{K} \alpha_i)} \) is the multivariate Beta function.  \( \Gamma(\alpha_i) \) is the Gamma function, which generalizes the factorial function.

In the context of binary categorization, the parameters of the Dirichlet distribution are correlated with the belief distribution, and the Dirichlet distribution can be viewed as the conjugate prior to the categorization distribution. 
Specifically, when the Dirichlet distribution is used as a prior for the categorical distribution, its parameters can be expressed as prior belief measures for different categories. The prior beliefs are then combined with the input data using Bayes' theorem to compute a posterior belief distribution. This posterior belief distribution can be further used to compute the confidence and overall prediction uncertainty for each category. Thus, the parameters of the Dirichlet distribution play a key role in the belief distribution, allowing the model to combine a priori knowledge and data to make inferences for more accurate and reliable categorization results.

In order to model uncertainty, the parameters of the Dirichlet distribution need to be determined. Our theoretical framework allows for the use of evidence collected from the data to obtain belief distributions. Evidence refers to the indicators obtained from the inputs to support categorization, and is closely related to the parameters of the Dirichlet distribution. According to Dempster-Shafer Evidence Theory (DST) \cite{dempster2008upper,fidon2024dempster}, in the K-categorization problem, the model attempts to assign a belief distribution to each category and an overall uncertainty of the entire framework. Thus, for each input, there are K+1 non-negative belief distribution values that sum to 1, as shown in Eq. \ref{eq5.1}.

\begin{equation}\label{eq5.1} u+\sum_{k=1}^{K}b_{k}=1  \end{equation}

where $u$ and $b_{k}$ denote the overall uncertainty and the probability of the $k^{th}$ class, respectively.

For each input, associate the parameters of the Dirichlet distribution $\alpha=[\alpha_1,\cdots,\alpha_K]$ with the evidence $e=[e_{1},\cdots,e_{K}]$. Specifically, $e_{K}$ determines the parameter $\alpha_K$ of the Dirichlet distribution, i.e., $\alpha_K=e_{K}+1$. Then, the belief quality $b_k$ and uncertainty measure $u$  are computed as follows:
\begin{equation}\label{eq5.2} b_k=\frac{e_k}{S}=\frac{\alpha_k-1}{S} \end{equation}
\begin{equation}u=\frac{K}{S} \end{equation}
where $S=\sum_{i=1}^{K}(e_i+1)=\sum_{i=1}^{K}\alpha_i$ is the strength of the Dirichlet distribution, which can be thought of as the total amount of evidence. Eq.\ref{eq5.2} describes the phenomenon that as the amount of evidence for the $K^{th}$ category increases, the probability of the $K^{th}$ category increases; conversely, as the total amount of evidence observed decreases, the total uncertainty increases.

Opinion consists of the predicted probabilities \( p_k \) for each category and the decision uncertainty, i.e., \( Opinion=\{\{p_k\}_{k=1}^{K},u\}\). In the fake audio detection task, these correspond to the decisions "Unknown", "Bonafide", and "Spoofed", respectively.  \( p_k \) is the mean  of the corresponding Dirichlet distribution and is computed as:

\begin{equation}
    p_k = \frac{\alpha_k}{\sum_{j=1}^{K} \alpha_j}
\end{equation}

The evidence is obtained by a deep neural network known as the evidential network. It is obtained by deforming a designed neural network. This is done by replacing functions whose output may be negative in the last layer of the neural network with a non-negative function. 
The evidential network is different from the traditional deep neural network classifier. First, while the output of a traditional neural network classifier is a single score indicating the predicted probability of the corresponding label, our model uses a Dirichlet distribution to parameterize the probability of each predicted probability, thus enabling the modeling of the second-order probability and uncertainty of the output. Second, traditional neural network classifiers typically use a softmax function for classification, but such output confidence tends to lead to over-confidence in the neural network model. Our model avoids this problem by adding an overall uncertainty measure. Some past methods\cite{2015Dropout, 2016Simple}usually require additional computation during inference to output uncertainty, but since uncertainty can only be obtained in the inference stage, it is difficult to train models with both high accuracy and robustness and reasonable uncertainty within a unified framework. As a result, the limitations of these algorithms (e.g., the inability to obtain uncertainty directly) also limit the utility of plausible classification. On the other hand, our model integrate uncertainty modeling in a unified framework that allows for seamless training of models and calculation of uncertainty, which contributes significantly to the utility of plausible classification.

\subsection{Learning to Generate Evidence}
In this section, we will discuss how to train a neural network to obtain evidence and then use it to obtain the parameters of the Dirichlet distribution. According to the study \cite{2018Efficient}, neural networks are capable of extracting evidence from inputs to support classification decisions, and thus traditional neural network-based classifiers can be transformed into evidence-based classifiers with minor changes. Specifically, a traditional neural network classifier can be transformed into an evidence-based classifier by replacing its softmax layer with a non-negative activation function layer. Doing so ensures that the network outputs non-negative values, which are regarded as evidence vectors, and thus the parameters of the Dirichlet distribution can be obtained.

For traditional neural network-based classifiers, cross-entropy loss is usually used:
\begin{equation}\label{eq5.4}L_{ce}=-\sum_{j=1}^{K}y_{ij}log(p_{ij}) \end{equation}
where $p_{ij}$ is the predicted probability of the $i^{th}$ sample of the $j^{th}$ class. 
For the model in this chapter, given the evidence for the $i^{th}$ sample obtained through the evidential neural network, the parameters of the Dirichlet distribution $\alpha_i$ (i.e., $\alpha_i=e_i+1$) can be obtained to form the evidence. After a simple modification of Eq. \ref{eq5.4}, the adjusted cross-entropy loss can be obtained:
\begin{displaymath}\label{eq5.51}  L_{ace}(\alpha_i)=
	\int[\sum_{j=1}^{K}-y_{ij}log(p_{ij})]\frac{1}{B(\alpha_i)}\prod\limits_{j=1}^{K}p_{ij}^{\alpha_{ij}-1}dp_i 
  \end{displaymath}
  \begin{equation}\label{eq5.5}  =
	\sum_{j=1}^{K}y_{ij}(\psi(S_i)-\psi(\alpha_{ij}))  
  \end{equation}
where $\psi(\cdot)$ denotes the digamma function and the Eq. \ref{eq5.5} is the integral of the cross-entropy loss function determined by $\alpha_i$. While the loss function described above ensures that correct labels for each sample produce more evidence than other classes of labels, it does not guarantee that incorrect labels produce less evidence. Therefore, it is desired that the evidence for incorrect labels in the model be progressively scaled down to close to 0. To this end, the following KL scatter term is introduced:
\begin{displaymath}\label{eq5.61} KL[D(p_i|\tilde{\alpha}_i)||D(p_i|1)]=
	log(\frac{\Gamma(\sum_{k=1}^{K}\tilde{\alpha}_{ik})}{\Gamma(K)\prod_{k=1}^{K}\Gamma(\tilde{\alpha}_{ik})})  \end{displaymath}
 
 \begin{equation}\label{eq5.6} +\sum_{k=1}^{K}(\tilde{\alpha}_{ik}-1)[\psi(\tilde{\alpha}_{ik})-\psi(\sum_{j=1}^{K}\tilde{\alpha}_{ij})]  \end{equation}
where $\tilde{\alpha}_{i}=y_i+(1-y_i)\bigodot\alpha_{i}$ is the Dirichlet distribution-adjusted parameter that avoids the evidence of correct labeling to be zero, and $\Gamma(\cdot)$ is the gamma function.

Thus, given the parameters $\alpha_{i}$ of the Dirichlet distribution for each sample $i$, the loss of specificity for that sample is:
\begin{equation}\label{eq5.7}L(\alpha_{i})=L_{ace}(\alpha_i)+\lambda_tKL[D(p_i|\tilde{\alpha}_i)||D(p_i|1)] \end{equation}
where $\lambda_t>0$ is the balancing factor. In the experiment, $\lambda_t$ can be gradually increased as the training progresses to prevent the network from focusing too much on the KL scatter term in the initial stage of training, which may otherwise result in the network not being able to optimize the parameters well enough to output a uniform distribution.

\section{Experiments}

\begin{figure*}[ht!]
    \centering
    \includegraphics[width=1\linewidth]{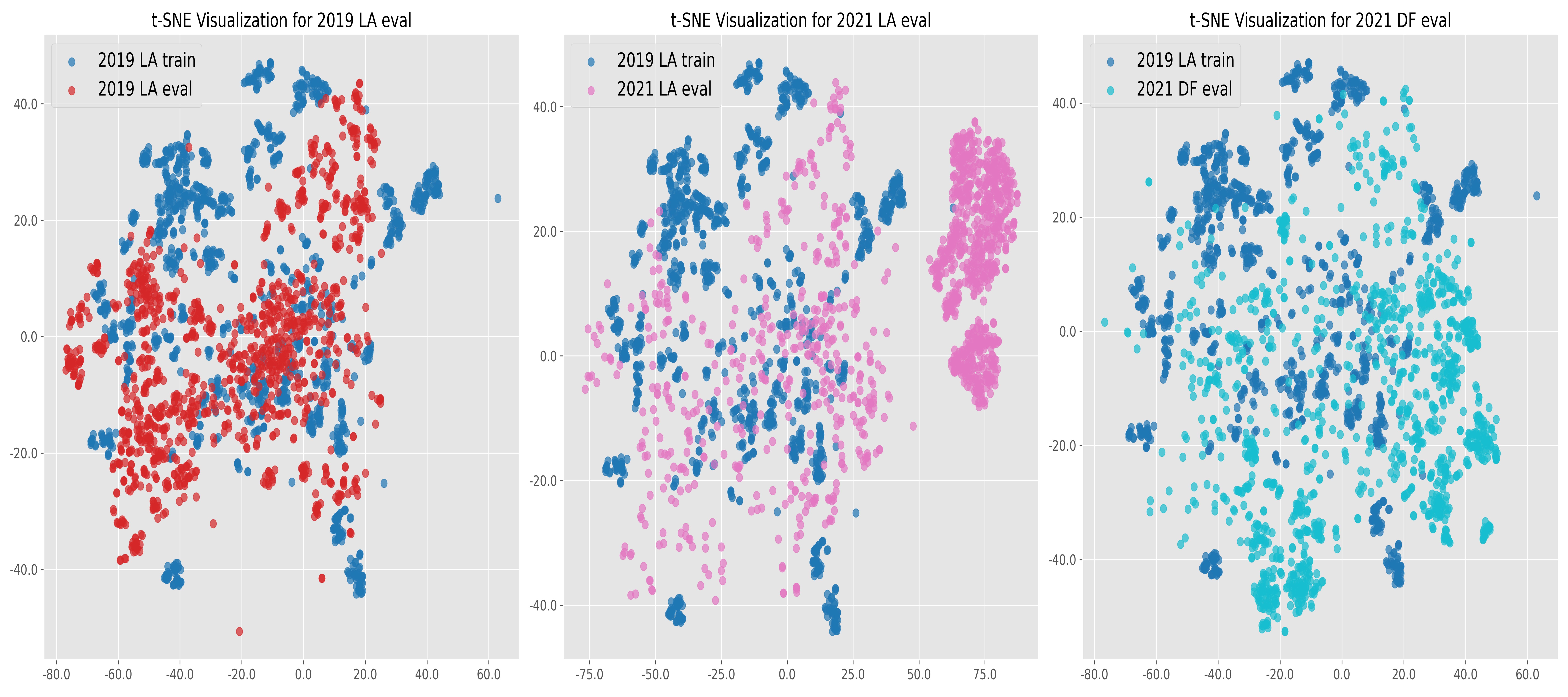}
    \caption{ASVspoof datasets t-SNE Visualization}
    \label{fig:t-sne}
\end{figure*}

\subsection{Datasets}
The ASVspoof series datasets stand as meticulously designed datasets tailored for the investigation of anti-spoofing measures in automated speaker verification.
 In our experiments, we utilize the training and development sets of the ASVspoof 2019 LA task dataset to train our model. And the proposed models are evaluated on the ASVspoof 2019 logical access (LA) task \cite{todisco2019asvspoof}, ASVspoof 2021 LA task \cite{yamagishi2021asvspoof}, and ASVspoof 2021 deepfake (DF) task \cite{yamagishi2021asvspoof}. 

\textbf{The ASVspoof 2019 LA dataset} consists of both bonafide utterances and spoofed utterances generated by 19 different spoofing attack algorithms. The training and development sets include six attack types (A01–A06), while the evaluation set introduces 13 additional attack strategies (A07–A19). All data samples are pristine and free from additional noise. To better understand the dataset’s characteristics, we visualized the Mel spectrogram features of all datasets using the t-SNE method, as shown in Figure \ref{fig:t-sne}. The visualization indicates that the 2019 LA training dataset has a similar feature distribution with the 2019 LA evaluation dataset. Among all the evaluation datasets, the detection model faces the least difficulty with this dataset.

\textbf{The ASVspoof 2021 LA evaluation dataset} expands on the 2019 LA dataset, containing 181,566 utterances. Unlike its predecessor, this dataset incorporates noise by transmitting each speech sample through various telephone systems, including Voice over IP (VoIP) and the Public Switched Telephone Network (PSTN). Consequently, the 2021 LA dataset exhibits a significantly different feature distribution from the training dataset, as shown in Figure \ref{fig:t-sne}.

\textbf{The ASVspoof 2021 DF evaluation dataset} consists of utterances from multiple sources, covering a range of spoofing attack strategies and artifacts introduced by different codecs. Both bonafide and spoofed speech samples in the DF track undergo processing with various lossy codecs. As depicted in Figure \ref{fig:t-sne}, this dataset displays diverse and distinct feature distributions compared to the training dataset, making it the most challenging among all the evaluation datasets.

\subsection{Evaluation Metrics}
The ASVspoof datasets offer two effective assessment measures. The evaluate metrics are equal error rate (EER) and minimum tandem detection cost function (t-DCF)\cite{tdct}. The lower the EER and t-DCF values, the higher the accuracy and reliability of the system.

In addition to evaluating the reliability of the model, we also examine its uncertainty estimation capabilities. Typically, the Expected Calibration Error (ECE)\cite{guo2017calibration} is widely used to measure the calibration of the model, which refers to the consistency between the predicted probabilities and the actual accuracy. However, considering that the uncertainty distribution is uneven, we adopt an improved metric, the adaptive Expected Calibration Error (aECE)\cite{nixon2019measuring}, for evaluation. This calibration is optimal when the metric is 0.

aECE adaptively groups predictions into R bins based on confidence, with each bin containing an equal number of predictions but varying widths. The error between the predicted confidence and the actual accuracy of each bin is then computed.

\begin{equation}
    \mathrm{aECE}=\frac{1}{R} \sum_{r=1}^{R}\left|{conf}\left(b_{r}\right)-{acc}\left(b_{r}\right)\right|,
\end{equation}

where, $b_r$ denotes the $r^{th}$ bin. $conf(b_r)$ denotes the average prediction confidence of the $b_r$, and $acc(b_r)$ denotes the actual accuracy of the $b_r$.

To evaluate the gap between model calibration and ideal calibration, we designed a quantitative metric called Prediction Confidence Consistency (PCC) as an auxiliary indicator. A smaller value of PCC indicates better calibration performance.
\begin{equation}
    \mathrm{PCC}=\sum_{r=1}^{R}\left|\frac{{conf}\left(b_{r}\right)}{{acc}\left(b_{r}\right)}-1\right|,
\end{equation}

\begin{table}[!t]
    \centering
    \scriptsize 
    \setlength{\tabcolsep}{1.2mm} 
    \caption{The  EER comparison on the ASVspoof series dataset}
    \begin{tabular}{cccccccccccccccc}
        \toprule
        Model  & 2019 LA & 2021 LA & 2021 DF \\
        \midrule
        CQCC-GMM(Baseline)\cite{todisco2019asvspoof}\cite{yamagishi2021asvspoof} & 9.57 &15.62 &25.56
        \\
        LFCC-GMM(Baseline)\cite{todisco2019asvspoof}\cite{yamagishi2021asvspoof} & 8.09 &19.30 &25.25
        \\
        LFCC-LCNN(Baseline)\cite{yamagishi2021asvspoof} & -&9.26  &23.48
        \\
        RawNet2(Baseline)\cite{yamagishi2021asvspoof} & -& 9.50 & 22.38
        \\
         \midrule
        Res-TSDNet\cite{hua2021towards} & 1.64& -&-  \\
        
        CQT+SE-Res2Net50 \cite{li2021replay} & 2.50& -&-   \\
        LFCC+LCNN-Dual attention \cite{ma2021improved}  & 2.76&- &-   \\
        LFCC+Resnet18-AM-Softmax [56]   & 3.26 & &  \\
        LFB+ResNet18-GAT-T \cite{tak2021graph} & 4.71&-  &-  \\
        LFB+ResNet18-GAT-S \cite{tak2021graph}  & 4.48&-  &-  \\
        
        LFCC+Siamese CNN \cite{lei2020siamese}  & 3.79& - &-  \\
        LFCC+DARTS \cite{wang2022fully}     &4.82 &-    &-          \\
        Wav2vec+DARTS \cite{wang2022fully}  &2.18  &-   &-         \\
        \midrule
        lightweight TDNN CE\cite{caceres2021biometric}  &-  &19.20  &- \\
        
        MFM-thin-ASSERT34\cite{wen2022multi} &-  & 17.35 &-   \\
        MFM-ASSERT18 \cite{wen2022multi}   &-   &  17.41 &-   \\
        GMM-MobileNet \cite{wen2022multi}  &-   &  8.75 &  20.08 \\
        
        SE-ResNet18 \cite{kang2021crim}    &-   &-   &23.13  \\
        WaveletCNN \cite{liu2023leveraging} &-   &-   &24.41  \\
        SE-Rawformer\cite{fathan2022mel}    &-   &-   &21.65  \\
        \midrule
        AASIST* & 1.52 & 8.16   &20.28
        \\
        \textbf{Trusted AASIST} & \textbf{1.33} & \underline{\textbf{7.65}}    &\textbf{19.91}
        \\
        RawNet2* & 5.37 &9.01   &24.67
        \\
       \textbf{Trusted RawNet2} & \textbf{4.55} & \textbf{8.11}   &\textbf{22.38}
        \\
        RawGAT-ST* & 1.52 &  12.42& 20.74
        \\
        \textbf{Trusted RawGAT-ST} & \underline{\textbf{1.29}} & \textbf{9.96 }& \underline{\textbf{17.70}}
        \\
        \bottomrule
        \multicolumn{4}{l}{* Represents the reproduced system.}
    \end{tabular}
    \label{Tab:asvspoof}
\end{table}

\begin{table*}[!ht]
    \centering
    \scriptsize 
    \setlength{\tabcolsep}{1.2mm} 
    \caption{The performance comparison on the ASV2019  LA dataset.}
    \begin{tabular}{cccccccccccccccc}
        \toprule
        Model  & A07 & A08 & A09 & A10 & A11 & A12 & A13 & A14 & A15 & A16 & A17 & A18 & A19 & Pooled EER\% / t-DCF \\
        \midrule

        AASIST*      & 0.30 & \textbf{0.30} & \textbf{0} & 0.54 & \textbf{0.17} & 0.38 & 0.13 & 0.14 &\textbf{ 0.34} & 1.36 & 2.52 & 4.53 & \textbf{0.99} & 1.52 / 0.0424
        \\
        Trusted AAIST &\textbf{ 0.26} & 0.39 & 0.04 & \textbf{0.42}& 0.26 & \textbf{0.34} & \textbf{0.08} & \textbf{0.12} & 0.38 & \textbf{1.03} &\textbf{ 1.97} & \textbf{4.17} & 1.16 & \textbf{1.33} / \textbf{0.0408}
        \\
        \midrule
        
        RawNet2*  &\textbf{0.26} & 4.86 & 0.22 & \textbf{0.36} &\textbf{ 0.32} & 0.51 & 0.24 & \textbf{0.26} & \textbf{0.30} & \textbf{0.59} & 10.49 & 17.07 & 1.99 &  5.37/ 0.1323
        \\
        Trusted RawNet2 & 0.34 &\textbf{3.96} & \textbf{0.20} & 0.43  & 0.34 & \textbf{0.39} & \textbf{0.22 }& 0.30 & 0.39& 0.66& \textbf{7.57} & \textbf{14.20} & \textbf{1.78} &  \textbf{4.55} / \textbf{0.1112}
        \\
        \midrule
        RawGAT-ST*      & 1.14 & \textbf{0.50} & \textbf{0.02} & 1.36 & 0.26 & 1.58 & 0.17 & 0.30 & 1.14 & 1.18 & 2.29& \textbf{3.96} & \textbf{0.83} & 1.52 / 0.0496
        \\
        Trusted RawGAT-ST &\textbf{0.52} & 1.03 & 0.04 & \textbf{0.61}& \textbf{0.24} & \textbf{0.79} & \textbf{0.06} & \textbf{0.10} & \textbf{0.52} & \textbf{ 0.83} &\textbf{ 2.17} & 4.47 & 1.05 & \textbf{ 1.29} / \textbf{0.0352}
        \\
        \bottomrule
        \multicolumn{4}{l}{* Represents the reproduced system.}
    \end{tabular}
    \label{Tab:2019}
\end{table*}

\begin{table*}[!ht]
    
    \centering
    \scriptsize 
    \setlength{\tabcolsep}{0.8mm} 
    \caption{The performance comparison on the ASV2021  LA dataset.}
    \begin{tabular}{cccccccccccccccc}
        \toprule
        Model  & A07 & A08 & A09 & A10 & A11 & A12 & A13 & A14 & A15 & A16 & A17 & A18 & A19 & Pooled EER\% / t-DCF \\
        \midrule
        AASIST*  & 6.74 & 7.39 & 3.46 & 7.71 & 6.33 & \textbf{7.75} &  5.70 & 5.80 & 7.35& 8.00 & 12.51 & 15.95 & \textbf{7.59} & 8.16 / 0.4149 \\
        Trusted AAIST  & \textbf{6.30} &\textbf{5.88} & \textbf{0.95} & \textbf{7.57} & \textbf{4.40} &  8.53 & \textbf{2.17} & \textbf{2.68} & \textbf{7.03} & \textbf{7.43} & \textbf{9.30} & \textbf{15.65} & 9.50 & \textbf{7.65} / \textbf{0.3114} \\
        \midrule
        RawNet2*  & \textbf{1.71 }&7.59 &1.83 &2.00&2.53 &2.85 &\textbf{1.19} & 3.14 & 2.52 & 3.11 & \textbf{20.59} &27.93 & 6.52 &   9.01/ 0.3616 \\
        Trusted RawNet2 & 1.87 &\textbf{6.36} & \textbf{1.52} & \textbf{1.89} & \textbf{1.80} & \textbf{2.44} &1.26 & \textbf{2.60} & \textbf{2.15} & \textbf{2.55} & 20.88 & \textbf{23.67} & \textbf{5.24}  & \textbf{8.11} / \textbf{0.3486}
        \\
        \midrule
        RawGAT-ST*  & 14.79 & \textbf{6.87} & \textbf{4.03} & 17.01 & 7.89 & 16.36 & 9.43 & \textbf{6.65} & 13.39 & 10.14 &\textbf{ 13.32 }& \textbf{18.29} & \textbf{9.62} & 12.42/ 0.5220
        \\
        Trusted RawGAT-ST & \textbf{9.46} & 9.40 & 4.62 & \textbf{9.68} & \textbf{5.68} & \textbf{10.41} & \textbf{4.94} & 6.75 &\textbf{9.07} & \textbf{7.79} & 14.47 & 19.22 & 10.32 & \textbf{9.96} / \textbf{0.4633}
        \\
        \bottomrule
        \multicolumn{4}{l}{* Represents the reproduced system.}
    \end{tabular}
    \label{Tab:2021}
\end{table*}

\begin{table}[!h]
    
    \centering
    \scriptsize 
    \setlength{\tabcolsep}{1.2mm} 
    \caption{The performance comparison on the ASV2021  DF dataset.}
    \begin{tabular}{ccccccc}
        \toprule
        Model  & T& W.C. & NAR & NnAR & U & Pooled EER\% 
        \\
        \midrule
        AASIST*  &13.41  & \textbf{14.66} &  25.99  &23.61 &21.55 &20.28
        \\
        Trusted AAIST  & \textbf{12.31} &15.82 &\textbf{25.60} &\textbf{23.10 }&\textbf{20.70} &\textbf{19.91}
        \\
        \midrule
        RawNet2*  & 23.11  &22.55    &25.85 &25.60  &23.33 &24.67   
        \\
        Trusted RawNet2 & \textbf{17.89} &\textbf{18.88} & \textbf{25.28} & \textbf{25.18} & \textbf{19.68}  & \textbf{22.38}
        \\
        \midrule
        RawGAT-ST*  & 14.45 & 17.82 & 25.62 & 23.08 & 20.24 & 20.74  
        \\
        Trusted RawGAT-ST & \textbf{12.31} & \textbf{10.02} &\textbf{ 23.19} & \textbf{19.81} & \textbf{17.11} & \textbf{17.70} 
        \\
        \bottomrule
        \multicolumn{4}{l}{* Represents the reproduced system.}
    \end{tabular}
    \label{Tab:2021DF}
\end{table}

\subsection{Experimental Setup}
In our experiments, we select three advanced and representative models, AASIST \cite{jung2022aasist}, RawNet2 \cite{tak2021rawnet2}, and RawGAT-ST \cite{tak2021end}, as evidential networks and conduct performance assessments on the 3 tasks mentioned above.
AASIST has demonstrated state-of-the-art (SOTA) performance on several tasks in the ASVspoof 2019 dataset. To utilize it as the evidential network, we add a softplus layer after the final fully connected layer of AASIST to ensure the output is positive. 
RawNet2 is the baseline model for the ASVspoof 2021 challenge, which has performed well on several related tasks. Similarly, we made a slight modification by replacing the final layer log-softmax with softplus to avoid negative outputs. 
RawGAT-ST employs graph attention networks to capture patterns in both the time and frequency domains. By leveraging model-level fusion, it integrates temporal and spectral information, effectively improving detection performance. This approach surpasses the RawNet2 model on the ASVspoof 2019 LA evaluation set by learning the correlations between different audio segments. We also add a softplus layer after the final fully connected layer.
The proposed method can transform the basic models into trusted models(i.e., trusted AA-SIST, trusted Rawnet2, and trusted RawGAT-ST). 

For the training phase of trusted models, the hyperparameters are maintained in accordance with the baseline settings without alteration.

\subsection{Results on ASVspoof 2019 LA task}

The ASVspoof 2019 LA evaluation dataset contains 108,978  utterances generated using 13 different methods (A07-A19). Table \ref{Tab:asvspoof} compares our trusted models with the state-of-the-art methods and two baseline systems on the ASVspoof 2019 LA task. 
On the ASVspoof 2019 LA task, the EER metrics of the trusted models reach 1.33 \%, 4.55 \%, and 1.29 \% respectively. The results in Table \ref{Tab:asvspoof} demonstrate that trusted RawGAT-ST outperforms all other models in the EER metric.
Table \ref{Tab:2019} shows the performance comparison of the basic models and the proposed trusted models. Compared to the AASIST model, the trusted AASIST model achieves a reduction in EER and min t-DCF by  12.5 \% and  3.7 \%, respectively. Similarly, the trusted RawNet2 model shows a decrease in EER and min t-DCF by  15.3 \% and  15.9 \%. The trusted RawGAR-ST model shows a decrease in EER and min t-DCF by  15.1 \% and  29.0 \%. 
Observing the performance of the models in each spoofing mode (A07-A019), we find that when the original model performs extremely well for a particular mode, our method suppresses the model's performance in the mode. For example, the EER of the trusted AASIST increases from 0 \% to 0.04 \% on A09, and the EER of the trusted RawGAT-ST increases from 0.02 \% to 0.04 \% on A09.

\begin{figure*}[!ht]
    \centering
    \includegraphics[width=0.99\linewidth]{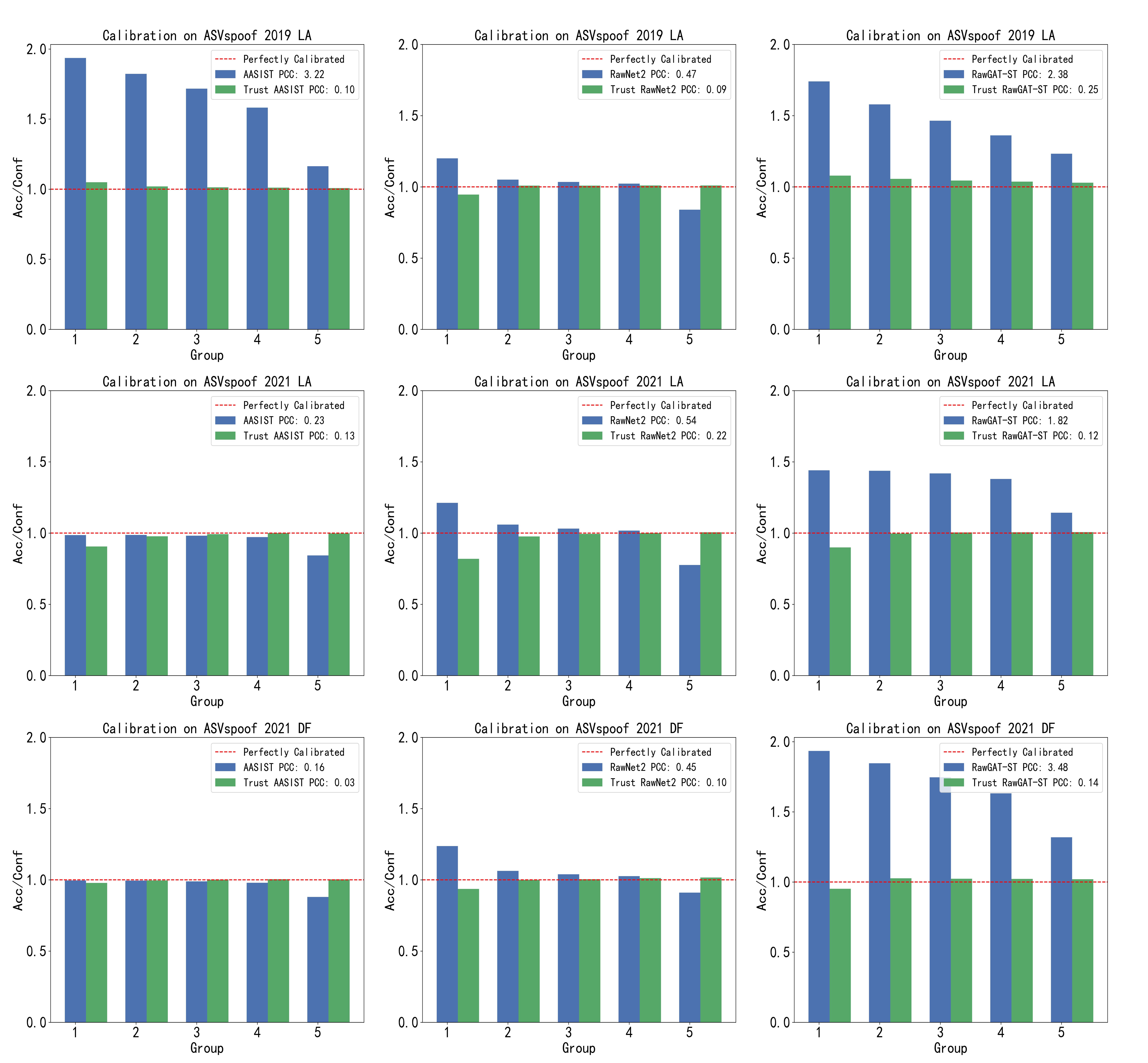}
    \caption{Calibration for different models on ASVspoof datastes }
    \label{fig:PCC}
\end{figure*}

\subsection{Results on ASVspoof 2021 LA task}

The ASVspoof 2021 LA evaluation dataset contains 181,566 audio samples. Moreover, compared to the ASVspoof 2019 LA evaluation dataset, it contains additional noise.
Table \ref{Tab:asvspoof} compares the trusted models with the existed models and four baseline systems on the ASVspoof 2021 LA task. On the ASVspoof 2021 LA task, the EER metrics of the trusted models reach 7.65 \%, 8.11 \%, and 9.96 \% respectively. The results in Table \ref{Tab:asvspoof} demonstrate that trusted AASIST outperforms all other models in the EER metric.
Table \ref{Tab:2021} shows the performance comparison between the original models and the trusted models on the dataset. Compared to the original AASIST model, the trusted AASIST model reduces the EER and min t-DCF by  6.2 \% and  24.9 \%, respectively. Besides, compared to the original RawNet2 model, the trusted RawNet2 model achieves a reduction in EER and min t-DCF by  9.9 \% and  3.6 \%. Compared to the original RawGAT-ST model, the trusted RawGAT-ST model achieves a reduction in EER and min t-DCF by  19.8 \% and  11.2 \%. 
Observing the performance of the models in each spoofing mode (A07-A019), We find similar inhibitory effects. For instance, the EER of the trusted RawNet2 increases from 1.19 \% to 1.26 \% on A13.

\subsection{Results on ASVspoof 2021 DF task}

The ASVspoof 2021 DF evaluation dataset contains 611,829 audio samples, which are processed with different lossy codecs. 
Table \ref{Tab:asvspoof} compares the trusted models with the existed models and four baseline systems on the ASVspoof 2021 DF task.
On the ASVspoof 2021 DF task, the EER metrics of the trusted models reach 19.91 \%, 22.38 \%, and 17.70 \% respectively. The results in Table \ref{Tab:asvspoof} demonstrate that trusted RawGAT-ST outperforms all other models in the EER metric. 
Table \ref{Tab:2021DF} shows the performance comparison between the basic models and the trusted models on this dataset . Compared to the basic AASIST model, the trusted AASIST model reduces the EER  by  1.8 \%. Similarly, compared to the basic models, the trusted RawNet2 and RawGAT-ST models achieve reductions in EER by  9.3 \% and 14.6 \%, respectively. 
The models perform relatively consistently in each spoofing mode on the ASVspoof 2021 DF task, and our method shows stable optimization effects.

\begin{table}[t]
    
    \centering
    \scriptsize 
    \setlength{\tabcolsep}{1.2mm} 
    \caption{The aECE comparison of models.}
    \begin{tabular}{ccccc}
        \toprule
        Model  & 19 LA & 21 LA & 21 DF & Avg aECE
        \\
        \midrule
AASIST & 0.371 & 0.046 & 0.032 & 0.150 \\
Trusted AASIST & \textbf{0.019} & \textbf{0.024} & \textbf{0.006} & \textbf{0.016} \\
\midrule
RawNet2 & 0.086 & 0.098 & 0.079 & 0.088 \\
Trusted RawNet2 & \textbf{0.017} & \textbf{0.040} & \textbf{0.018} & \textbf{0.025} \\
\midrule
RawGAT-ST & 0.309 & 0.251 & 0.390 & 0.317 \\
Trusted RawGAT-ST & \textbf{0.046} &\textbf{ 0.022} & \textbf{0.026} & \textbf{0.031} \\
        \bottomrule
    \end{tabular}
    \label{Tab:aECE}
\end{table}

\subsection{Improvement in Uncertainty Estimation}
To demonstrate the improvement in uncertainty estimation before and after applying the method, we conduct comparison experiments that calculated the adaptive Expected Calibration Error (aECE) of the model before and after applying the method. 
It is important to note that the original detection model does not provide uncertainty estimates or confidence for its decisions. To facilitate the comparison of changes in uncertainty estimates, we normalized the model’s output scores to the range [0,1], interpreting them as decision confidence. This normalized confidence is then used to calculate the aECE.
Table \ref{Tab:aECE} shows that the aECE values of the trusted models are almost always lower than those of the regular models, indicating that the trusted models perform better and are more reliable in terms of uncertainty estimation. 
The average aEERs of the trusted models across multiple datasets are 0.016, 0.025, and 0.031. Compared to the original model, these represent a relative reduction of 89.3 \%, 71.6 \%, and 90.2 \%, respectively. The results in Table \ref{Tab:aECE} indicate that the trusted AASIST achieves the best aECE improvement in the ASVspoof 2019 LA task, reaching 94.9 \%.
Additionally, to more intuitively demonstrate the gap between model calibration and the ideal scenario, we group the data and calculate the ratio of accuracy to prediction confidence for each group, as shown in Figure 3. Theoretically, the closer this ratio is to 1, the better the calibration performance. Furthermore, we calculate the PCC to measure the distance between each model calibration and perfect calibration.
Experimental results show that the trusted models have more accurate uncertainty estimates.

\begin{figure}[t]
	\centering
	\includegraphics[scale=0.35]{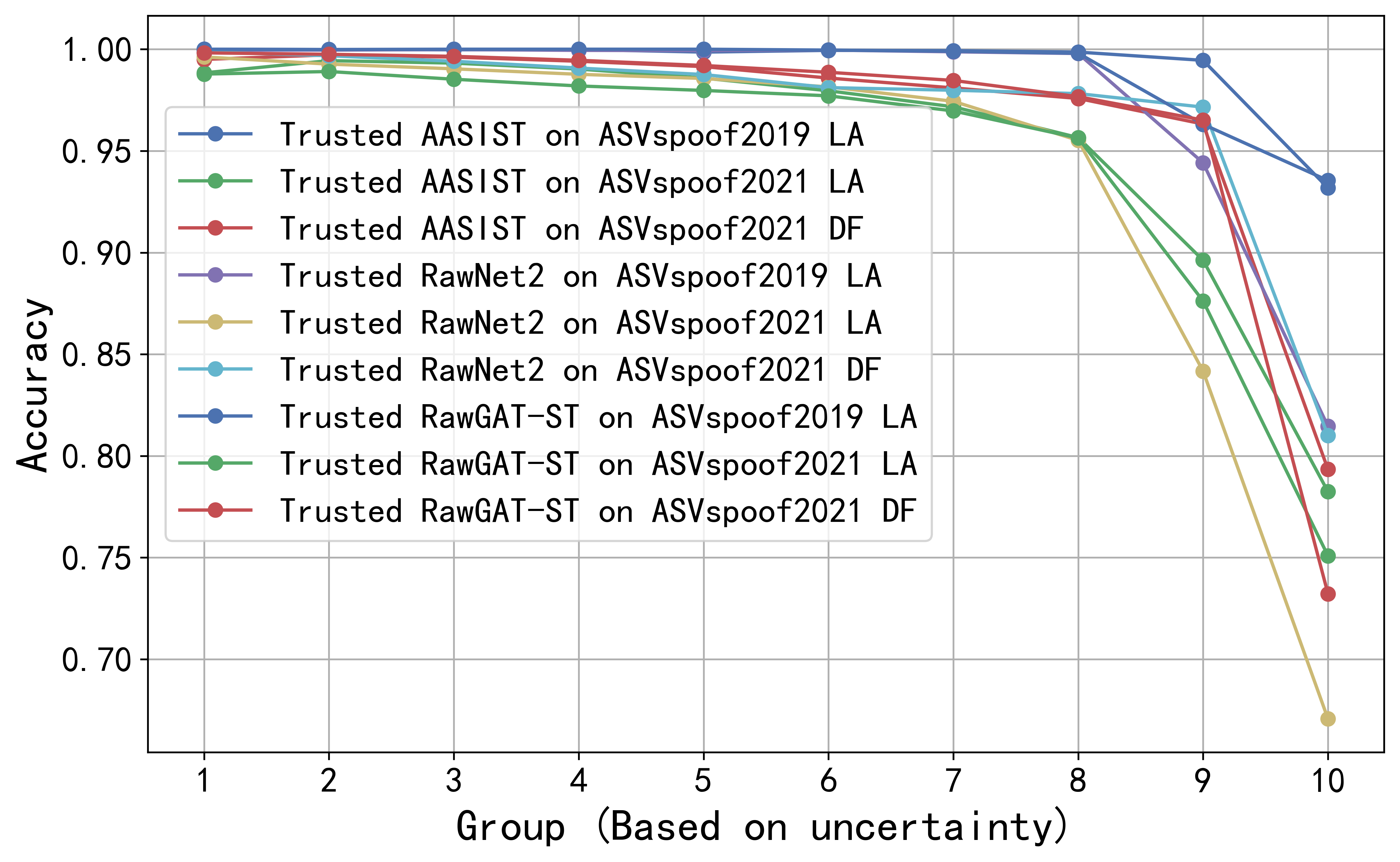}
	\caption{Based on uncertainty $u$, the samples are divided into 10 groups with the same number. The ordinate represents the accuracy of each group.}
 \label{f2}
\end{figure}

\subsection{Relationship between the Accuracy and Uncertainty}

Figure \ref{f2} visually illustrates the relationship between the uncertainty of decisions made by various trusted models and their prediction accuracy. As demonstrated in Figure \ref{f2}, the accuracy of classification decisions decreases with increasing uncertainty. When the model is confident in its judgment, its accuracy tends to be higher than 95 \%. When the model exhibits underconfidence in its predictions, its accuracy deteriorates significantly.
The results demonstrate that the trusted model can effectively flag uncertain predictions, indicating a higher level of reliability.

\section{Conclusion}
In this paper, we propose a trusted fake audio detection method based on Dirichlet distribution. The method is structured around three core stages: the generation of evidence, uncertainty modeling, and opinion generation. 
To be specific, evidential neural networks underpin evidence generation, while the Dirichlet distribution determined by evidence is used to model belief distribution. Then decision uncertainty and predictive probabilities of each categories form an opinion. Experimental validation firmly substantiates the efficacy of the proposed credible model. In comparison with state-of-the-art DNN-based techniques, the trusted model demonstrates a better performance. 
Experimental results show that our approach achieves significant improvements in EER, min t-DCF , aECE, and PCC metrics compared to the existed advanced models on the ASVspoof series datasets. 
Additionally, the relationship between the uncertainty provided by the model and the accuracy of the classification indicates that the proposed model can assess the uncertainty of its decisions during the inference phase effectively, further enhancing its reliability.












\bibliographystyle{elsarticle-num}  
\bibliography{arvix}  

\end{document}